\documentclass[twocolumn,aps,prl,reprint,superscriptaddress,showpacs,longbibliography,footnoteinbib]{revtex4-2}
\usepackage{lmodern}
\usepackage[T1]{fontenc}
\usepackage{float}
\usepackage{amsmath}
\usepackage{amssymb}
\usepackage{graphicx}
\usepackage[unicode=true,
 bookmarks=true,bookmarksnumbered=false,bookmarksopen=false,
 breaklinks=false,pdfborder={0 0 1},backref=false,colorlinks=false]
 {hyperref}
\hypersetup{pdftitle={Title}}

\makeatletter

\newcommand{\lyxdot}{.}

\usepackage{units}\usepackage{wasysym}

\usepackage{bm}

\@ifundefined{showcaptionsetup}{}{%
 \PassOptionsToPackage{caption=false}{subfig}}
\usepackage{subfig}
\makeatother

\begin{document}
\global\long\def\vect#1{\overrightarrow{\mathbf{#1}}}%

\global\long\def\abs#1{\left|#1\right|}%

\global\long\def\av#1{\left\langle #1\right\rangle }%

\global\long\def\ket#1{\left|#1\right\rangle }%

\global\long\def\bra#1{\left\langle #1\right|}%

\global\long\def\tensorproduct{\otimes}%

\global\long\def\braket#1#2{\left\langle #1\mid#2\right\rangle }%

\global\long\def\omv{\overrightarrow{\Omega}}%

\global\long\def\inf{\infty}%

\title{Topological phase transitions for any taste in 2D quasiperiodic systems}
\author{Tiago S. Gonçalves$^{1}$, Miguel Gonçalves$^{2}$, Pedro Ribeiro$^{2,3}$,
Eduardo V. Castro$^{1,2,3}$, Bruno Amorim$^{1}$}
\affiliation{Centro de F\'{ı}sica das Universidades do Minho e Porto, Departamento
de F\'{ı}sica e Astronomia, Faculdade de Ciências, Universidade do
Porto, 4169-007 Porto, Portugal}
\affiliation{CeFEMA, Instituto Superior Técnico, Universidade de Lisboa,
Av. Rovisco Pais, 1049-001 Lisboa, Portugal}
\affiliation{Beijing Computational Science Research Center, Beijing 100084,
China}
\begin{abstract}
In this paper we explore the effects of  quasiperiodicity in paradigmatic
models of Chern insulators. We identify a plethora of topological
phase transitions and characterize them based on spectral and localization
properties, by using a wealth of exact numerical methods. Contrary
to uncorrelated disorder, gap closing and reopening topological transitions
can be induced by quasiperiodicity. These can separate widely different
phases, including (i) trivial and Chern insulators, both with ballistic
states near the gap edges; (ii) Chern insulators with critical states
around the gap edges or (iii) Chern and trivial insulators respectively
with ballistic and localized gap-edge states. Transition (i) is of
the same nature as clean-limit topological transitions due to the
ballistic character of the gap-edge states, but at the same time resembles
(quasi-)disorder driven topological Anderson insulator phenomena.
On the other hand, transitions (ii) and (iii) have no clean-limit
counterpart. Additionally, quasiperiodicity can also induce topological
transitions into a trivial state for which the gap closes and does
not reopen, a scenario that resembles more what is observed with uncorrelated
disorder. However, we found that such transitions can also be non-conventional
in that they can be accompanied by the emergence of intermediate metallic
and critical phases where the Chern number is not quantized. Our results
show that a rich variety of topological phase transitions, not previously
realized experimentally nor predicted theoretically can be attained
when applying quasiperiodic modulations to simple models of Chern
insulators. Such models have previously been realized experimentally
in widely different platforms, including in optical lattices and photonic
or acoustic media, where quasiperiodicity effects can be incorporated.
The novel topological phase transitions here unveiled can therefore
in principle be observed experimentally with state-of-the-art techniques. 

\end{abstract}
\maketitle

\section{Introduction}

Topological insulators are currently a hot topic of research due to
their unusual properties when compared to trivial, common band insulators
\citep{RevModPhys.82.3045,QZrmp11,bernevigBook,Chiu2016}. After the
discovery of the quantum Hall effect \citep{Klitzing1980} and its
theoretical explanation using concepts of topology \citep{TKNN82,NTW85},
the quantum anomalous Hall effect, where a topological phase can be
induced without the need of an applied homogeneous magnetic field,
was proposed by Haldane \citep{Haldane1988}. This was later realized
experimentally on several systems under zero net magnetic field \citep{Chang2013,Jotzu2014,Checkelsky2014,Chang2015}.
This makes quantum anomalous Hall insulators, or Chern insulators
according to the topological classification \citep{Chiu2016}, highly
appealing systems to study quantum topological matter at the fundamental
level, and to explore possible applications of their distinctive feature
-- topologically protected gapless surface states.

Topological systems are known to be robust to effects of uncorrelated
disorder \citep{Xiao2010} as long as these do not break any fundamental
symmetry. In the case of quantum Hall systems and anomalous quantum
Hall insulators, disorder is even crucial for the observation of quantized
Hall conductivity, since it localizes every state except for those
responsible for the quantized Hall current \citep{KM93,OMN+03,nagaosaQSHloc07}.
A well quantized Hall plateau is then the consequence of the Fermi
level laying in the gap (filled with localized states) which separates
the extended states that carry the topological invariant. Increasing
disorder strength leads, generally, the Chern insulator to a trivial
phase \citep{prodanBernevig,Prodan2011}, under the standard mechanism
referred to as ``levitation and annihilation'' of extended states
\citep{nagaosaQSHloc07}, although exceptions have been identified
\citep{Castro2015,Castro2016}. Interestingly, topological Anderson
insulator phases have also been discovered, in which disorder is responsible
to drive a phase transition from trivial to topological phases \citep{Shen2009,Groth2009,Song2012,Garcia2015,Orth2016}. 

A different class of systems that break translational invariance,
where even more exotic localization properties can occur, are quasiperiodic
systems (QPS). These systems have received a lot of attention due
to their non-trivial localization properties in one \citep{AubryAndre,Roati2008,Lahini2009,Schreiber842,Luschen2018}
and higher \citep{Huang2016a,PhysRevLett.120.207604,Park2018,PhysRevX.7.041047,PhysRevB.100.144202,Fu2020,PhysRevB.101.235121,Wang2020,Goncalves_2022_2DMat}
dimensions. In generic QPS, there can be fully extended, localized
and critical phases even in 1D \citep{PhysRevB.43.13468,PhysRevLett.104.070601,PhysRevLett.113.236403,Liu2015,PhysRevB.91.235134,PhysRevLett.114.146601,anomScipost,GoncalvesRG2022,Goncalves_CriticalPhase},
while the same is not possible in general in the presence of uncorrelated
disorder. More recently, renewed attention has been drawn to QPS in
the context of Moiré systems and flatband physics \citep{Balents2020,Andrei2020},
and due to the possibility of being simulated in widely different
platforms, including optical lattices \citep{PhysRevA.75.063404,Roati2008,Modugno_2009,Schreiber842,Luschen2018,PhysRevLett.123.070405,PhysRevLett.125.060401,PhysRevLett.126.110401,PhysRevLett.126.040603,PhysRevLett.122.170403}
and photonic \citep{Lahini2009,Kraus2012,Verbin2013,PhysRevB.91.064201,Wang2020,https://doi.org/10.1002/adom.202001170,Wang2022}
or even phononic \citep{PhysRevLett.122.095501,Ni2019,PhysRevLett.125.224301,PhysRevApplied.13.014023,PhysRevX.11.011016,doi:10.1063/5.0013528}
media. QPS are also very well-known for their intriguing topological
properties \citep{Kraus2012,PhysRevLett.109.116404,Verbin2013,Zilberberg21}
that can be seen as originating from a construction using additional
(virtual) dimensions \citep{Steurer1997,Zilberberg21}

Even though the intrinsic topological properties of QPS have been
widely studied, the effects of quasiperiodicity in conventional topological
systems, such as Chern insulators, has been less explored \citep{Fu_2021},
with preprints only appearing very recently in this direction \citep{sacramento_QPS,cheng2022_december}.
Here, by studying the effect of quasiperiodic potentials on paradigmatic
models of Chern insulators, we unveil a miriad of possibilities for
the interplay between topology, localization and spectral properties.
We obtain the fate of the phase diagrams of the Haldane model and
Bernevig-Hughes-Zhang (BHZ) models at half-filling in the presence
of quasiperiodic potentials. Additionally, we identify gapped and
gapless regions in this new phase diagram and relate all these results
with the localization properties of the states at the Fermi level/near
the gap edge. We identify a plethora of phase transitions of different
nature that were hitherto unknown for two-dimensional systems subject
to uncorrelated disorder. Hence, we aim to show that quasi-disorder/quasiperiodicity
offers a rich playground to explore new and interesting phenomena
in topological systems that are alien to clean systems and systems
subjected to random disorder.

\begin{figure}
\includegraphics[width=0.9\columnwidth]{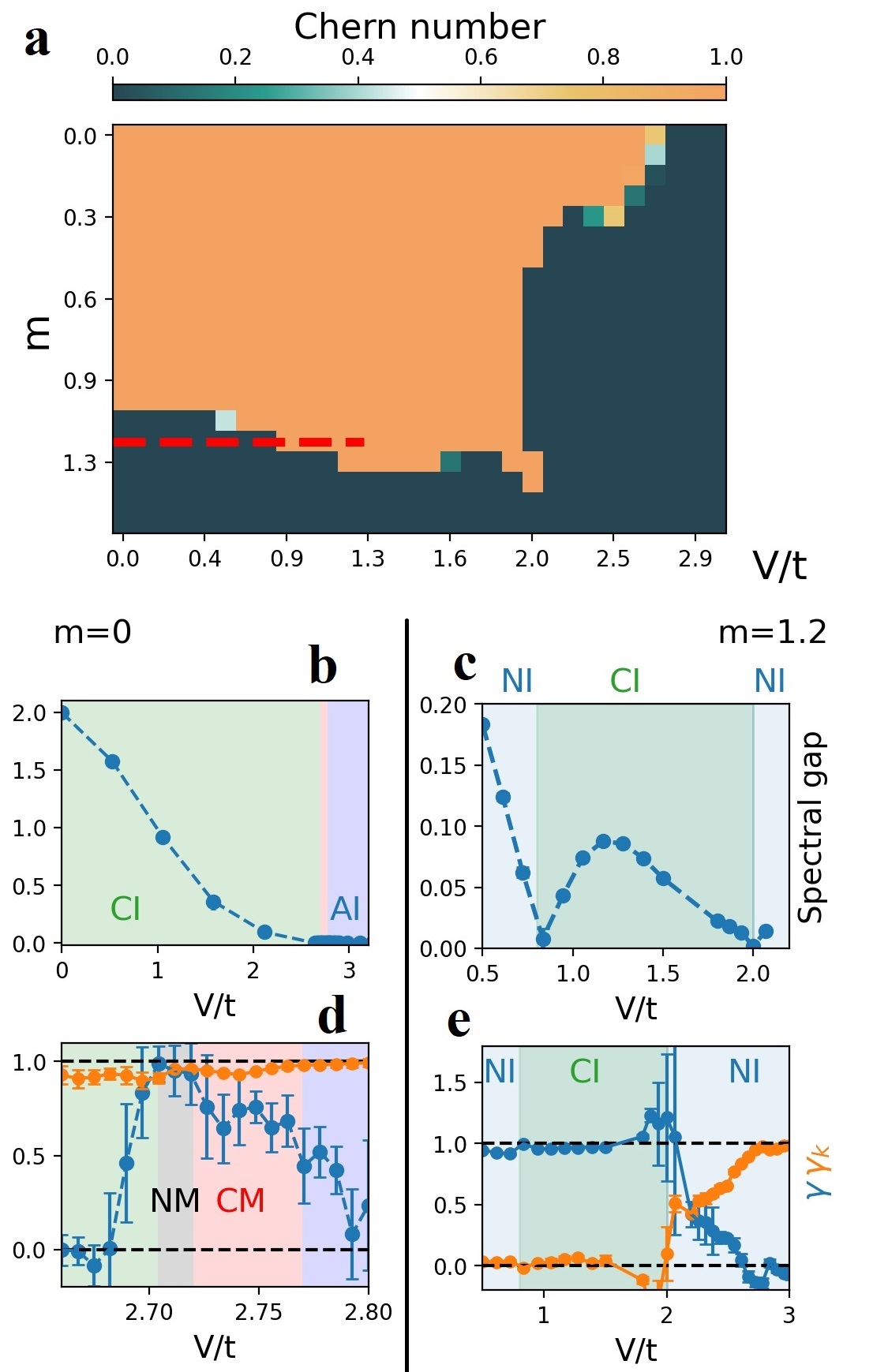}
\captionsetup{justification=raggedright,margin=0cm}
\caption{Summary of results for the Haldane model subjected to a quasiperiodic
potential of strength $V$, and with $t_{2}=0.2$ {[}see Eqs.$\,$\ref{eq:haldane_hamil},\ref{eq:disordered_haldane_hamil}{]}.
(a) Phase diagram of the topological phases for a system of size $N_{\text{u.c.}}=34\times34$
with an average over 50  configurations of random $\protect\vect r_{0}$
{[}Eq.$\,$\ref{eq:disordered_haldane_hamil}{]} and phase twists.
(b,c) Spectral gap for a system of size $N_{\text{u.c.}}=144\times144$,
respectively for $m=0$ and $m=1.2$.  We see that in the region where
the system is a topological insulator with $C=1$ the system has non-zero
gap. (d,e) Fractal dimension as a function of quasi-disordered potential
for $m=0$ and $m=1.2$. The different phases are labelled as Chern
insulator (CI), normal metal (NM), critical metal (CM), Anderson insulator
(AI) and gapped trivial/normal insulator (NI). \label{fig:haldane-summary}}
\end{figure}

\begin{figure}
\includegraphics[width=0.9\columnwidth]{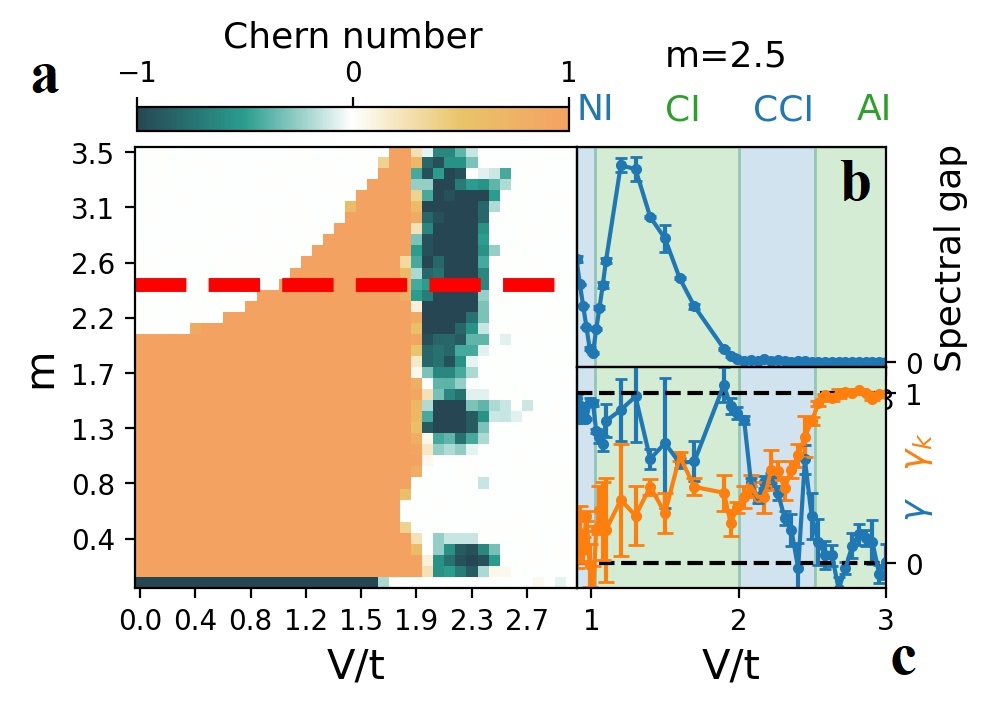}
\captionsetup{justification=raggedright,margin=0cm}
\caption{(a) Phase diagram of the topological phases of the BHZ model with
$m=2.5$. These results were obtained for a system of size $N_{\text{u.c.}}=34\times34$
with an average over 50 configurations. (b) Evolution of the spectral
gap as a function of quasi-periodic strength for $m=2.5$ {[}indicated
by red dashed line in (a){]}. . The different phases are labelled
as gapped trivial/normal insulator (NI), Chern insulator (CI), critical
Chern insulator (CCI) and Anderson insulator (AI). (c) Fractal dimension
as a function of quasi-periodic strength.\label{fig:summary-square}}
\end{figure}

In Fig.$\,$\ref{fig:haldane-summary} we provide a summary of the
most important results obtained for the Haldane model subject to an
incommensurate potential. Naturally, for large strengths of the potential
$V$, the topological phase of the original system is destroyed. For
$m=0$ we identify an interesting scenario where the Chern insulator
(CI) is first suppressed because the system enters gapless normal
(diffusive) and critical metallic phases (respectively NM and CM),
in which case the Chern number becomes non-quantized (but finite),
before it becomes a gapless Anderson insulator (AI) with null Chern
number at larger $V$ {[}Fig.$\,$\ref{fig:haldane-summary}(b){]}.
This is corroborated by the energy gap and real-space fractal dimension
results in Figs.$\,$\ref{fig:haldane-summary}(b,d) respectively
Additionally, we see the emergence of a CI phase with $C=1$ starting
from a trivial/normal insulator (NI) phase {[}indicated by the dashed
red line in Fig.$\,$\ref{fig:haldane-summary}(a){]}, accompanied
by the closing and reopening of the gap {[}see Fig.$\,$\ref{fig:haldane-summary}(c){]}.
This situation is analogous to the topological Anderson insulator,
with the important difference that the gap edge states in both phases
are ballistic (characterized by a vanishing momentum-space fractal
dimension $\gamma_{k}$). Such transition is therefore of the same
nature as topological transitions in the clean limit, which is not
possible under uncorrelated disorder {[}Fig.$\,$\ref{fig:haldane-summary}(c){]}.
For even higher potential strength, an additional gap closing and
reopening transition takes place into a trivial phase with localized
gap-edge states (characterized by a vanishing real-space fractal dimension
$\gamma$).  Adding a quasiperiodic potential to the BHZ model also
gives rise to interesting unconventional phase transitions. Examples
are shown in Fig.$\,$\ref{fig:summary-square}, where there is a
$V$-driven topological phase transition between two gapped {[}Fig.$\,$\ref{fig:summary-square}(b){]}
topological phases with symmetric Chern numbers (CI and CCI). The
gap edges within these phases contain critical states around the transition,
that are characterized by non-integer fractal dimensions {[}see Fig.$\,$\ref{fig:summary-square}(c){]}. 

The paper is organized as follows: In Sec.\,\ref{sec:method}, we
introduce the studied models and briefly explain the numerical methods
employed throughout the manuscript. In Sec.\,\ref{sec:results} we
carry out a detailed numerical analysis of the different possible
transitions, that includes the characterization of topological, spectral
and localization properties. . In Sec.\,\ref{sec:conclusions} the
key results are summarized their implications are discussed.


\section{Models and methods}

\label{sec:method}

We start by considering the Haldane model \citep{Haldane1988} with
Hamiltonian written as
\begin{align}
H_{0}=-t\sum_{\left\langle i,j\right\rangle }c_{i}^{\dagger}c_{j}+t_{2}\sum_{\left\langle \left\langle i,j\right\rangle \right\rangle }e^{i\phi_{i,j}}c_{i}^{\dagger}c_{j}+\eta\sum_{i}\zeta_{i}c_{i}^{\dagger}c_{i},+\text{H.c}.,\label{eq:haldane_hamil}
\end{align}
Here, $c_{i}^{\dagger}(c_{i})$ are creation (annihilation) operators
defined in the two triangular sub-lattices $A$ and $B$ that form
the honeycomb lattice. The first term in Eq.\,(\ref{eq:haldane_hamil})
corresponds to hopping between nearest neighbor sites $\langle i,j\rangle$,
and couples sublattices $A$ and $B$. The second term describes complex
next-to-nearest neighbor hopping between sites $\langle\langle i,j\rangle\rangle$,
with amplitude $t_{2}e^{i\phi_{ij}}$ and $\phi_{ij}=\nu_{ij}\phi$,
where $\nu_{ij}=(2/\sqrt{3})(\hat{\boldsymbol{d}}_{1}\times\hat{\boldsymbol{d}}_{2})=\pm1$
with $\hat{\boldsymbol{d}}_{1}$ and $\hat{\boldsymbol{d}}_{2}$ two
unit vectors along the two bonds connecting $\langle\langle i,j\rangle\rangle$.

In the following $t$ is set to unity and $t_{2}=0.2t$ which ensures
a direct gap with no band overlapping \citep{Haldane1988}. The third
term corresponds to a staggered potential, where $\zeta_{i}=1$ if
$i\in A$ and $\zeta_{i}=-1$ if $i\in B$. These last two terms are
respectively responsible for breaking time-reversal and inversion
symmetries and thus for opening non-trivial and trivial topological
gaps at the Dirac points, respectively. In the absence of disorder
the phase diagram of the Haldane model in the $(\eta,\phi)$ parameter
space encompasses a trivial phase with vanishing Chern number and
two topological non-trivial phases with $C=\pm1$ respectively. 

We also consideredthe BHZ, whose Hamiltonian in  reciprocal space
can be written as

\begin{equation}
H(\vect k)=\vect h(k)\cdot\vect{\sigma},\label{eq:band_ham_sq}
\end{equation}
Where $\vect{\sigma}$ is the Pauli matrices vector $\vect{\sigma}=\left(\sigma_{x},\sigma_{y},\sigma_{z}\right)$
and 

\[
\vect h(k)=\left(sin(k_{x}),sin(k_{y}),M-tcos(k_{x})-tcos(k_{y})\right).
\]

The system is gapped for every value of $M$ except $M=\left\{ 0,-2,2\right\} $,
where the topological phase transitions take place. In what follows,
we set $t=1$.

The quasiperiodic potential is introduced in both models through the
term 

\begin{equation}
\begin{aligned}V_{\text{qp}}\left(\vect r\right)=V\sum_{\vect b_{i}}\cos\left(\beta\left(\vect r-\vect r_{0}\right)\cdot\vect b_{i}\right).\end{aligned}
\label{eq:disordered_haldane_hamil}
\end{equation}

where $\beta$ is an irrational number that is responsible for the
incommensurate character of this potential, $\vect b_{i}$ are the
primitive vectors of the reciprocal lattice and $\vect r_{0}$ is
a shift in the center of the potential. In the case of the quasiperiodic
Haldane model we take this quasiperiodic potential to be the same
in both sublattices, however, in the case of the BHZ we take this
potential to take symmetric value in each sublattice.

In practice, we wish to obtain numerical results for increasingly
large unit cells of these quasiperiodic systems, and in order to do
this we need $\beta$ to have a series of rational and commensurate
approximants. To meet this condition in the best way possible, we
chose $\beta$ to be the golden ratio, $\varphi=\frac{1+\sqrt{5}}{2}$,
which can be approximated by the ratio of consecutive members of the
Fibonacci sequence, $f_{n+1}/f_{n}\rightarrow\varphi$. Additionally,
the numerics are obtained for different configurations of potential
center shift $\vect r_{0}$and the twists, $\theta$, of the twisted
boundary conditions, defined as

\[
\begin{aligned}\phi_{n}^{\theta}\left(x+L_{x},y\right) & \equiv\braket{x+L_{x},y}{\phi_{n}^{\theta}}=\phi_{n}^{\theta}\left(x,y\right)e^{i\theta_{x}}\\
\phi_{n}^{\theta}\left(x,y+L_{y}\right) & \equiv\braket{L_{x},y+L_{y}}{\phi_{n}^{\theta}}=\phi_{n}^{\theta}\left(x,y\right)e^{i\theta_{y}}
\end{aligned}
.
\]

In the following, we extend the phase diagram of the Haldane model
and the model on the square lattice by increasing the quasidisorder
strength until the topological phases are destroyed. We identify the
topological nature of each phase by computing the Chern number using
Fukui's method \citep{Fukui2005} as implemented in Ref.~\citep{Zhang2013},
a variant that is suitable to deal with systems with broken translational
invariance.

The localization properties are probed using inverse participation
ratios (IPR). To define them, we start by writing a given eigenstate
of the Hamiltonian $\ket{\phi_{n}}$ in the local site basis as $\ket{\phi_{n}}=\sum_{i}\phi_{\vect r_{i}}^{n}\ket{\vect r_{i}}$,
where $\ket{\vect r_{i}}=c_{i}^{\dagger}\ket 0$. Then, the real-space
IPR is defined as $\text{IPR}_{n}=\sum_{i}\abs{\phi_{\vect r_{i}}^{n}}^{4}$.
By carrying out a fourier transform for the real-space amplitudes
$\phi_{\vect r_{i}}^{n}$, we get momentum-space amplitudes through
which we can compute the momentum-space IPR - the $\text{IPR}_{k}$
- in the same way. The finite-size scaling behaviour of these quantities
provides a good metric to fully understand the localization properties
of a given state. In particular, we can study the $\textit{fractal dimension}$,
$\gamma$ ($\gamma_{k}$), defined as 

\[
\text{IPR}\sim\frac{1}{N_{\text{sites}}^{\gamma}}\quad\text{IPR}_{k}\sim\frac{1}{N_{\text{sites}}^{\gamma_{k}}},
\]
where $N_{\text{sites}}$ is the total number of sites in the system.
For ballistic (localized) states we have $\gamma=1;\gamma_{k}=0$
( $\gamma=0;\gamma_{k}=1$), while for diffusive (normal) metallic
states we have $\gamma=\gamma_{k}=1$. Finally, for critical states
the fractal dimensions take non-integer values either in real or momentum-space,
or in both. 

Finally, to study spectral properties of our system, namely the spectral
energy gap, we used both exact diagonalization or the the Kernel Polynomial
Method (KPM) as implemented in the KITE quantum transport software
\citep{Joo2020KITEHA}.

\section{Results}

\label{sec:results}

\subsection{Quasiperiodic Haldane model}

In this section discuss in detail the results obtained for the Haldane
model subject to an incommensurate potential, modeled in the fashion
described in the previous section. The topological phase diagram is
shown in Figure \ref{fig:haldane-summary}(a). We can see that for
small  $V$ the topological phase is robust. However, for sufficiently
high quasidisorder there is always a suppresion of the topological
phase. Another relevant feature os the presence of reentrant behaviour
indicated by the dashed red line, where we see an analogous situation
to the topological Anderson insulating phenomena observed for uncorrelated
disorder in the sense that a topological phase is induced by the presence
of quasidisorder.

In the following sections, we will study in detail two cases: the
quasi-disorder-driven phase transition (i) from a topological to a
trivial phase and (ii) from a trivial gapped phase to a topological
quasidisordered insulator and back to trivial insulator.

We describe the phases of our system by taking into account three
main characteristics: the spectral energy gap, the topological invariant,
and the localization of states near the gap edge. In order to minimize
finite-size effects, we extrapolated the spectral gap to the thermodynamic
limit, $N_{\text{sites}}\rightarrow\infty$, by making a linear regression
of the value of the gap as a function of $1/N_{\text{sites}}$ and
extracting the predicted gap size for $1/N_{\text{sites}}=0$.

\subsubsection{From a topological to a trivial phase\label{subsec:haldane-m=00003D0}}

\begin{figure}
\begin{centering}
\subfloat{\includegraphics[width=0.95\columnwidth]{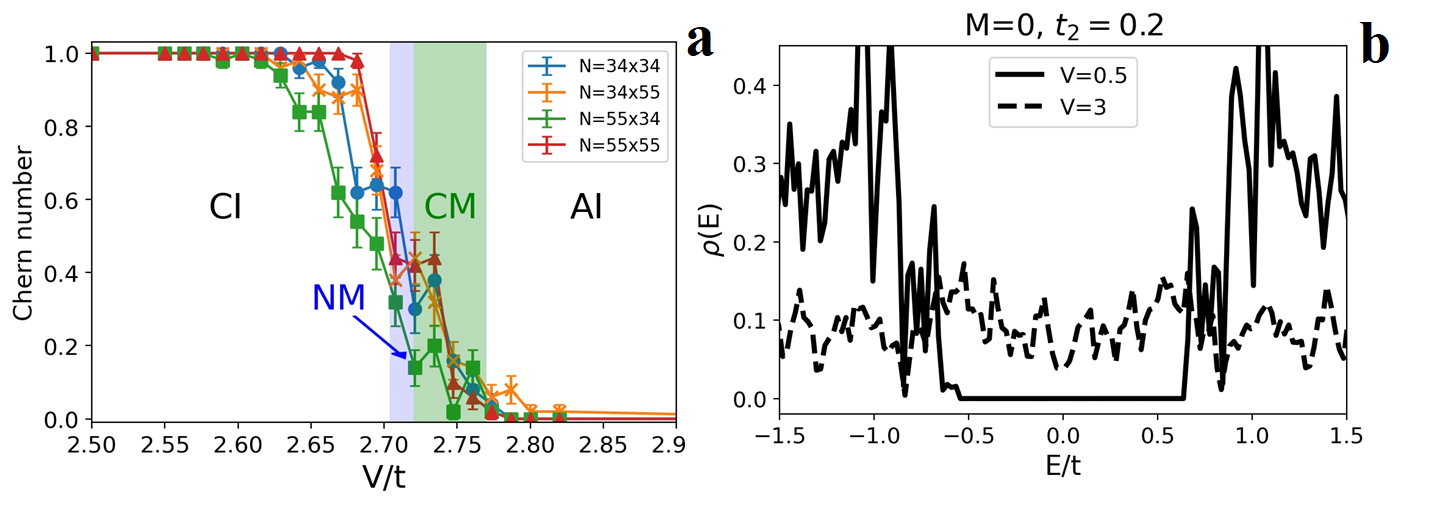}}
\par\end{centering}
\begin{centering}
\subfloat{\includegraphics[width=0.95\columnwidth]{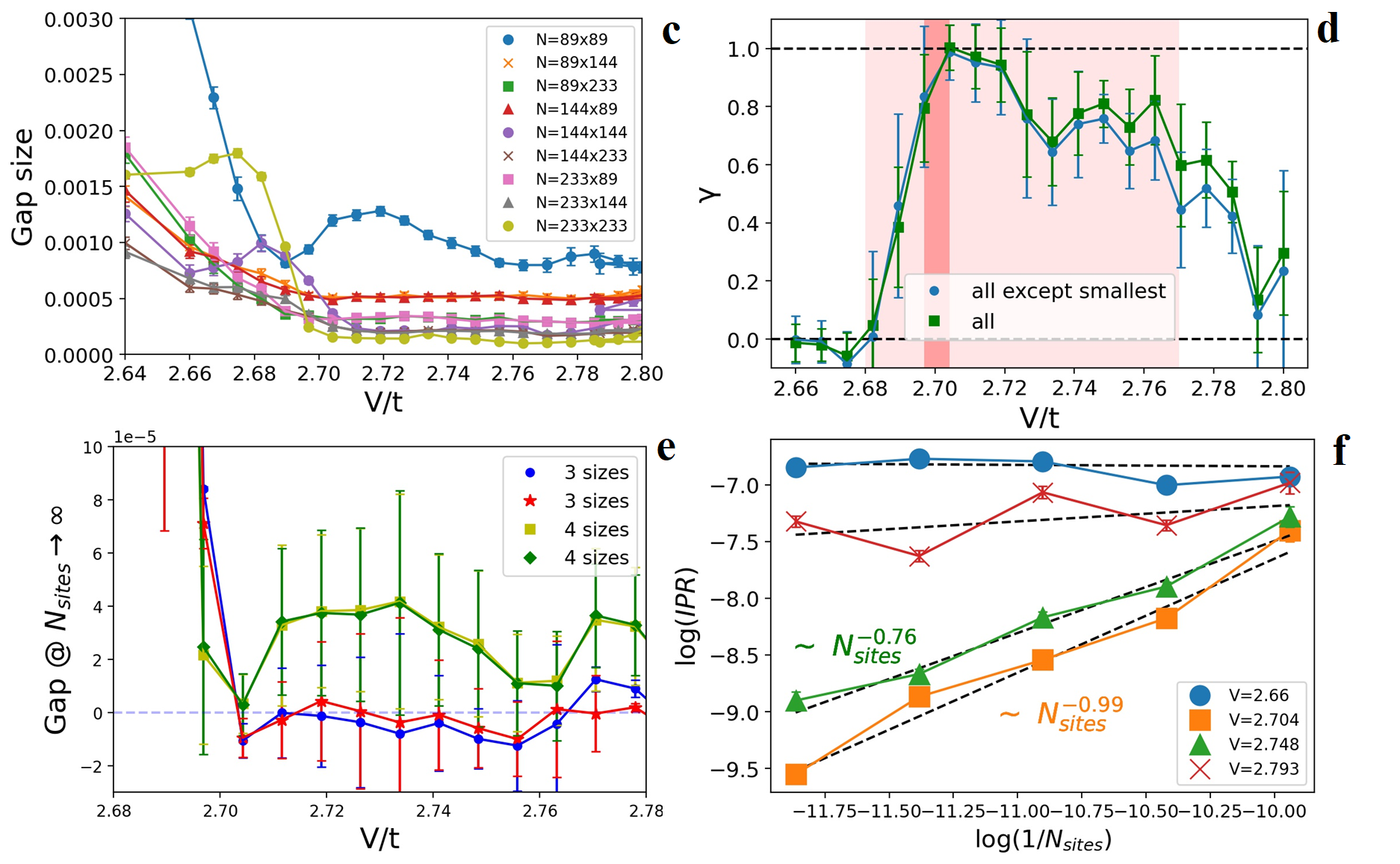}}
\par\end{centering}
\captionsetup{justification=raggedright,margin=0cm}
\caption{(a) Chern number as a function of incommensurate potential strength
$V/t$ for various system sizes averaged over 50 configurations. (b)
DoS of the Haldane model for $V=0.5$ (CI phase) and $V=3$ (AI phase)
obtained with KPM. (c) Evolution of the spectral gap for different
system sizes for the Haldane model with $m=0$. (d) Value of the spectral
gap in the limit $N_{\text{sites}}\rightarrow\infty$ obtain through
linear regression. The different curves represent different sets of
system sizes used to perform the linear regression. For 3 sizes we
used ($N_{\text{sites}}=233\times233,144\times233,144\times144$ and
$N_{\text{sites}}=233\times233,233\times144,144\times144$), and for
4 sizes we used ($N_{\text{sites}}=233\times233,144\times233,144\times144,233\times89$
and $N_{\text{sites}}=233\times233,233\times144,144\times144,89\times233$).
In (f) we explicitly represent a few the finite-size scalings of $\text{IPR}$
for a few selected values of quasi-disorder.\label{fig:haldane-m=00003D0-figs}}
\end{figure}

Figure \ref{fig:haldane-m=00003D0-figs}(a) shows a cut of the phase
diagram in Fig.$\,$\ref{fig:haldane-summary}(a) at $m=0$. Clearly,
for low values of quasidisorder we are in the presence of a Chern
insulator characterized by $C=1$. We can also see in Figs.$\,$\ref{fig:haldane-m=00003D0-figs}(b,c,e)
that in this region our system is gapped. Additionally, we can characterize
the localization properties of the states of the gap edge in this
region using the IPR in conjunction with the fractal dimension. These
results are shown in Fig.$\,$\ref{fig:haldane-m=00003D0-figs}(d),
indicating localized states, with fractal dimension compatible with
0. For $V\gtrsim2.7$, the system enters a gapless phase, as shown
in Fig.$\,$\ref{fig:haldane-m=00003D0-figs}(e), where the thermodynamic
limit extrapolation of the energy gap is compatible with $0$. At
sufficiently large $V$, the system enters a gapless topologically
trivial phase with non-quantized Chern number.  As shown in Fig.$\,$\ref{fig:haldane-m=00003D0-figs}(d),
this gapless region hosts a small normal (diffusive) metal (NM) region
with $\gamma\approx1$, and a critical region (CM) where the fractal
dimension exhibits an intermediate value of $\gamma\approx0.75$,
robust to increasing the system size (see also some explicit finite-size
scalings of the ${\rm IPR}$ in Fig.$\,$\ref{fig:haldane-m=00003D0-figs}(f),
from which $\gamma$ was extracted). Note, however, that in this case
we only observe real-space fractality since $\gamma_{k}\approx1$
as previously seen in Fig.$\,$\ref{fig:haldane-summary}(d). At even
larger $V$, the system remains gapless, but the Chern number becomes
exactly zero, which implies the presence of localized states at the
Fermi level. This is further corroborated by the decrease in $\gamma$
in this regime: we expect $\gamma\rightarrow0$ in the thermodynamic
limit. Thus, this phase is analogous to an Anderson insulator (AI)
observed with uncorrelated disorder, that in this case is instead
induced by the presence of quasi-disorder. What is notable about this
transition, however, is that it is mediated by a gapless region that
hosts diffusive or critical states at the Fermi level. These are evidenced
once again in Fig.$\,$\ref{fig:haldane-m=00003D0-figs}(d) where
we see, after a small normal (diffusive) metal region with $\gamma\approx1$,
a region where the fractal dimension exhibits an intermediate value
of $\gamma\approx0.7$, robust to increasing the system size. 

\subsubsection{From trivial to Chern insulator and back to trivial\label{subsec:From-trivial-to-quasidisordered}}

We now analyse in detail the situation where we start in a topologically
trivial phase and a Chern insulator phase is induced by quasi-disorder.
The Chern number calculations are shown in Fig.$\,$\ref{fig:gap-haldane-m=00003D1.2}(a).
In this case, unlike the one considered in the previous section, our
system is gapped for every value of $V$ except where the topological
transitions between the gapped normal insulators (NI) and Chern insulator
(CI) happen. This result is clearly evidenced in Fig.$\,$\ref{fig:gap-haldane-m=00003D1.2}(b),
where we plot the thermodynamic-limit extrapolations of the finite-size
gap calculations shown in Figs.$\,$\ref{fig:gap-haldane-m=00003D1.2}(c,d).

The ${\rm IPR}$ and ${\rm IPR}_{k}$ data used to compute the fractal
dimensions in Fig.$\,$\ref{fig:haldane-summary}(e) are shown in
Fig.$\,$\ref{fig:ipr_haldane_m=00003D1.2}. There, we see that the
${\rm IPR}_{k}$ becomes system-size independent in the small-$V$
NI phase and in the CI phase, signaling the presence of ballistic
gap-edge states in both. Upon increasing $V$, we have a transition
into a gapped insulator at $V=2.00\pm0.07$. At this large-$V$ NI
phase, however, the ${\rm IPR}$ becomes constant sufficiently away
from the transition, signaling the presence of localized gap-edge
states. As we decrease $V$ and approach the transition, the ${\rm IPR}$
starts to decrease with system size. Nonetheless, we expect the gap-edge
states to be localized in the large-$V$ NI phase, but with a localization
length that diverges at the transition - indeed we observe that the
${\rm IPR}$, that is inversely proportional to this localization
length, sharply increases with $V$ as we move away from the transition.
Since close to the transition the localization length can be very
large it is natural that we do not yet see the ${\rm IPR}$ converging
to a constant for the available system sizes, that are smaller than
this length. Therefore, in summary we unveiled two types of transitions
with the gap closing and reopening mechanism: one in which the gap-edge
states are ballistic, which is the same mechanism behind topological
transitions in clean systems; and another in which the gap edge states
change from ballistic to localized, with no counterpart in the clean
limit. 
\begin{center}
\begin{figure}[H]
\begin{centering}
\subfloat{\centering{}\includegraphics[width=0.95\columnwidth]{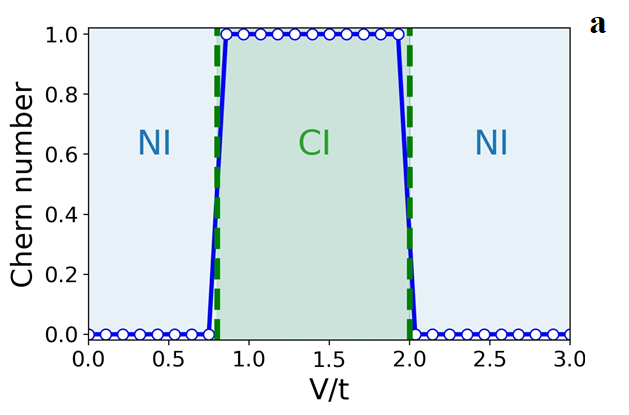}}
\par\end{centering}
\begin{centering}
\subfloat{\centering{}\includegraphics[width=0.9\columnwidth]{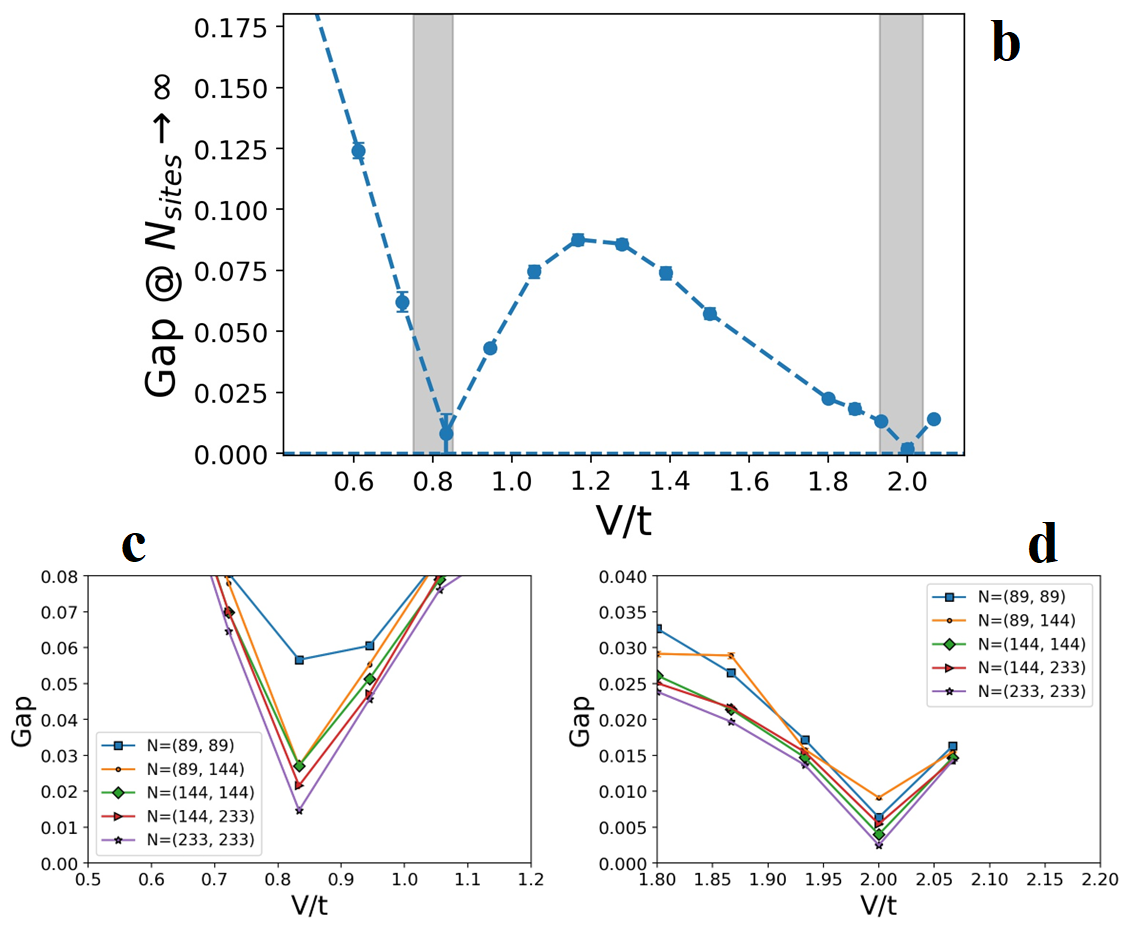}}
\par\end{centering}
\captionsetup{justification=raggedright,margin=0cm}
\caption{(a) Chern number results for the Haldane model with a staggered mass
$m=1.2$. We identify three distinct phases of our system controlled
by the amplitude of the quasi-periodic potential: a normal insulator
phase, a Chern insulator phase, and, finally, an Anderson insulator-like
phase. These results were obtained for a system with dimensions $34\times34$
averaged over 30 configurations. (b) Gap size in the limit $N_{\textit{sites}}\rightarrow\infty$
obtained through linear regression. The shaded grey areas represent
the regions where we expect to find the topological phase transitions.
(c) and (d) show a zoom around the first and second gap closing (respectively)
with different system sizes represented. \label{fig:gap-haldane-m=00003D1.2}}
\end{figure}
\par\end{center}

\begin{figure}
\begin{centering}
\subfloat{\centering{}\includegraphics[scale=0.55]{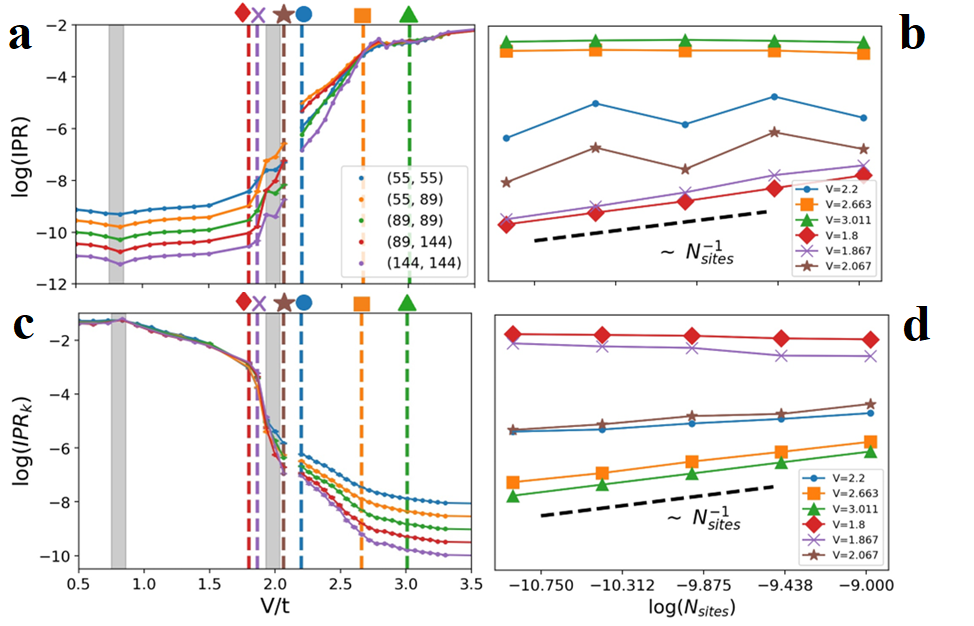}}
\par\end{centering}
\captionsetup{justification=raggedright,margin=0cm}
\caption{(a) Value of $\log\left(\text{IPR}\right)$ as a function of quasi-periodic
strength $V/t$ for different system sizes. The shaded regions indicate
where we saw the topological phase transitions. Once again, he shaded
grey areas represent the regions where we expect to find the topological
phase transitions in Figure \ref{fig:gap-haldane-m=00003D1.2}(a).(b)
$\log\left(\text{IPR}\right)\textit{vs}1/N_{\text{sites}}$ for a
few values of incommensurate potential strength. (c) and (d) show
the same as figures (a) and (b) but for the values of the $\text{IPR}_{k}$.
 \label{fig:ipr_haldane_m=00003D1.2}}
\end{figure}

\subsubsection{}

\subsection{Quasiperiodic BHZ Model}

In this section, we study the quasiperiodic BHZ described in $\ref{sec:method}$.
We have seen the topological phase diagram in Fig.$\,$\ref{fig:summary-square}(a).
Here we analyse the transitions crossed by the dashed red line in
this figure in more detail. Contrary to the Haldane model, increasing
quasi-disorder at fixed $m$ can induce two topological phases with
distinct Chern numbers and therefore a richer phase diagram. 

Setting $m=2.5$, we see that 3 distinct topological transitions occur:
(i) from a NI to a CI, (ii) from a CI to a distinct CI, and finally,
(iii) from a CI to an AI. Below, we once again characterize these
transitions using the same metrics used for the Haldane model. The
first transition is qualitatively identical to the transition described
in \ref{subsec:From-trivial-to-quasidisordered} for the Haldane model,
and thus we will not dwell on it again.

\subsubsection{Chern Insulator to Critical Chern Insulator}

\begin{figure}
\begin{centering}
\subfloat{\centering{}\includegraphics[width=0.9\columnwidth]{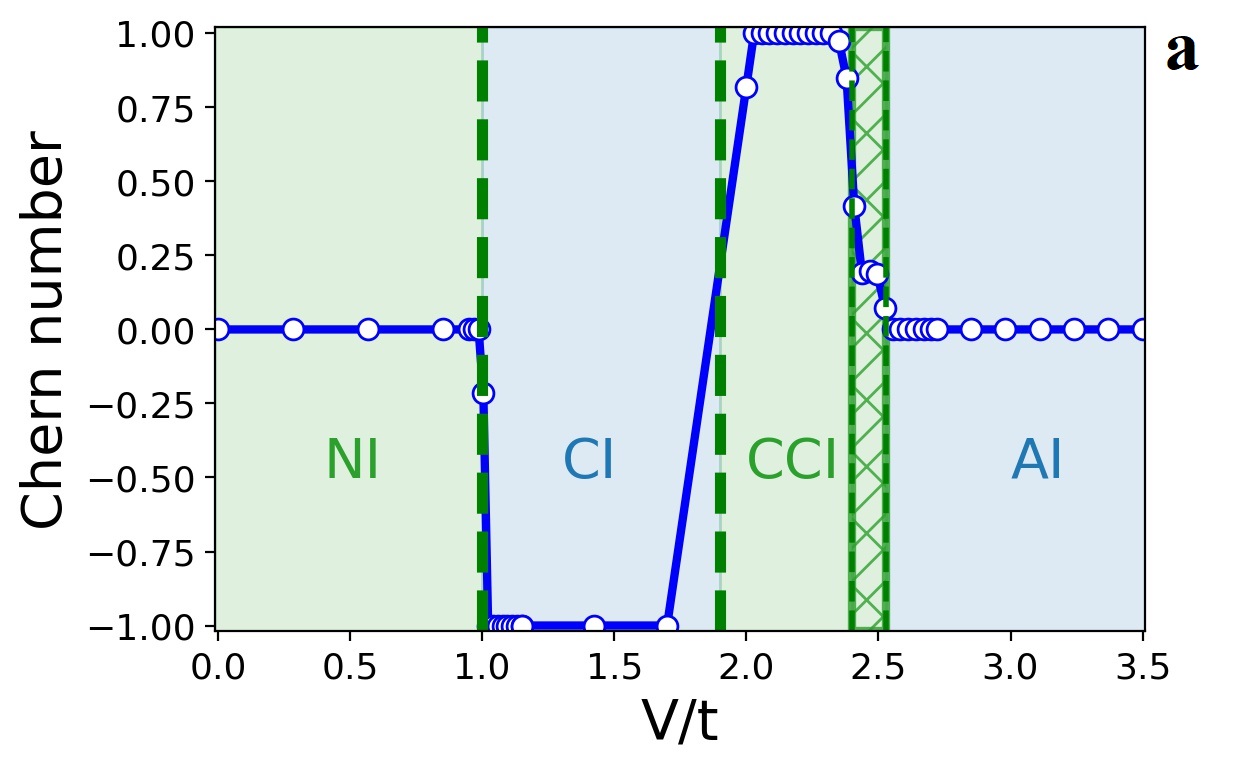}}
\par\end{centering}
\begin{centering}
\subfloat{\includegraphics[width=0.95\columnwidth]{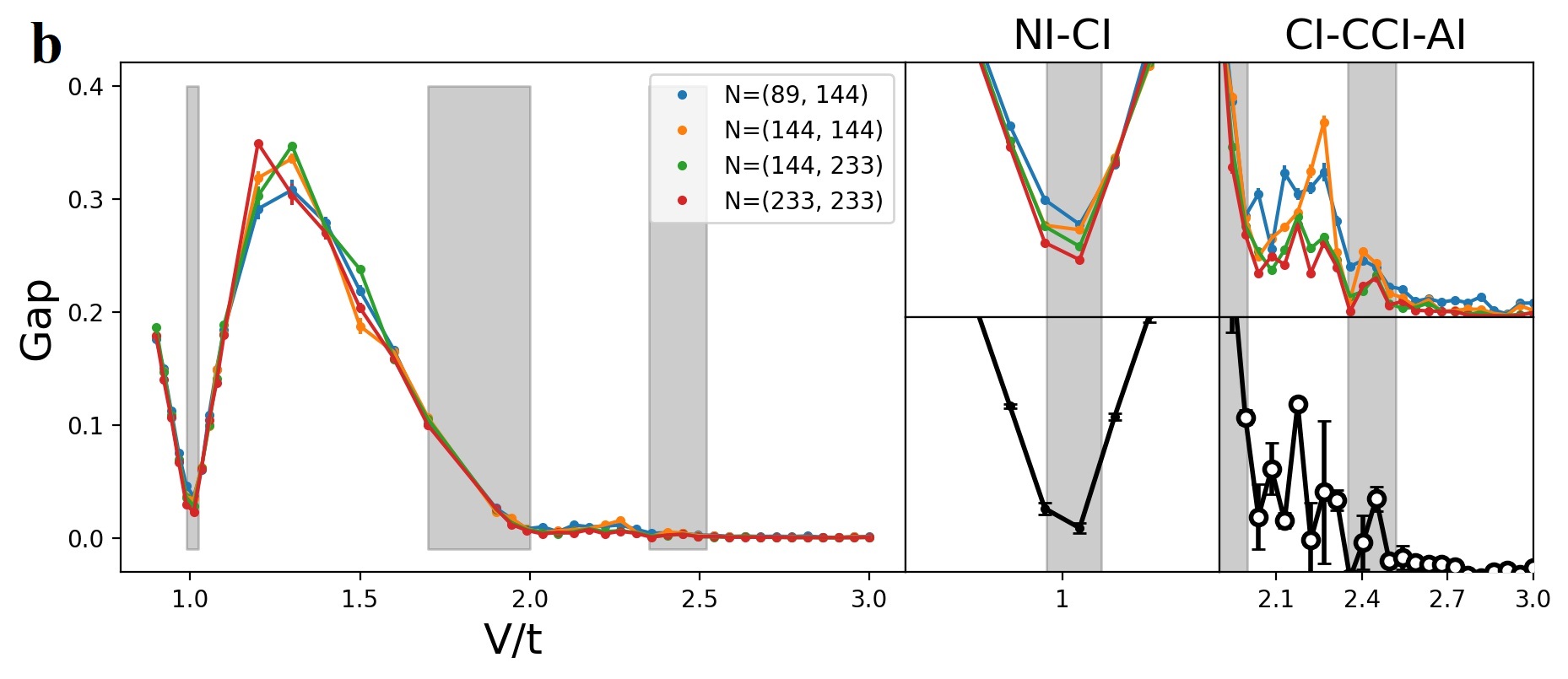}}
\par\end{centering}
\captionsetup{justification=raggedright,margin=0cm}
\caption{(a) Topological invariant as a function of quasi-periodic potential
strength. We observe three topological transitions at $V=\left(1.007\pm0.017\right)t$,$V=\left(1.95\pm0.15\right)t$,
and $V=\left(2.435\pm0.170\right)t$. The criss-crossed green region
denotes a range of value of $V/t$ where our results are not yet conclusive,
and is discussed in more detail in the body of the text. These results
are obtained for a system with size $N_{\text{sites}}=34\times34$
averaged over 200 configurations. (b) Evolution of the spectral gap
as a function of quasi-periodic disorder strength for different system
sizes. In the insets we show the two regions of interest with the
panels on top representing the values of the spectral for different
system sizes and the figures below showing the value in the limit
$N_{\text{sites}}\rightarrow\infty$ obtain through linear regression.
\label{fig:gap-square}}
\end{figure}

The second transition is from a $C=-1$ CI to a $C=1$CI as shown
in Fig.$\,$\ref{fig:gap-square}(a). The thermodynamic-limit extrapolation
of the gap in Fig.$\,$\ref{fig:gap-square}(b) suggest that this
is a gap closing and reopening transition since the system is gapped
in both phases. Even though no conclusive gap closing point is observed
accross the transition, the change in Chern number requires that this
must be the case and we just missed the precise point in the numerical
simulations.

For the $C=-1$ CI, the fractal dimension results in Fig.$\,$\ref{fig:summary-square}
are not very conclusive and suffer from severe finite-size effects,
as evidenced by the very large error bars. The reason for these large
errors can be seen in Fig.$\,$\ref{fig:IPR_square}, where no monotonic
scaling is observed for the IPR and ${\rm IPR}_{k}$ as a function
of system size. Nonetheless, close to the transition into the $C=1$
phase, our results strongly suggest that the gap-edge states are critical.
A clear example is shown in detail in Figs.$\,$\ref{fig:IPR_square}(g,h),
where we see that the size dependence of the ${\rm IPR}$ and ${\rm IPR}_{k}$
data for $V=1.6$ is very well described by a power-law with fractal
dimensions $\gamma\approx\gamma_{k}\approx0.6$. At the $C=1$ phase,
on the other hand, the gap-edge states are clearly critical, being
fractal both in real and momentum-space for the whole range of $V$
within this phase. This is unveiled by the clear scaling of both ${\rm IPR}$
and ${\rm IPR}_{k}$ in Fig.$\,$\ref{fig:IPR_square} yielding the
intermediate fractal dimensions shown in Fig.$\,$\ref{fig:summary-square}.
We therefore call this phase the critical Chern insulator (CCI). We
therefore have seen that on both sides of the CI-CCI transition, the
gap-edge states are critical, which is a novel feature compared to
uncorrelated disorder, for which the eigenstates only become critical
exactly at the transition.   Note that this phase should be distinguished
from the CM phase seen in subsection \ref{subsec:haldane-m=00003D0}
for the Haldane model, since in that case we were in the presence
of a gapless system and the Chern number was non-quantized. In the
CCI, the system is gapped and topological, but the gap-edge states
are critical.

\begin{center}
\begin{figure}[H]
\begin{centering}
\subfloat{\includegraphics[width=0.95\columnwidth]{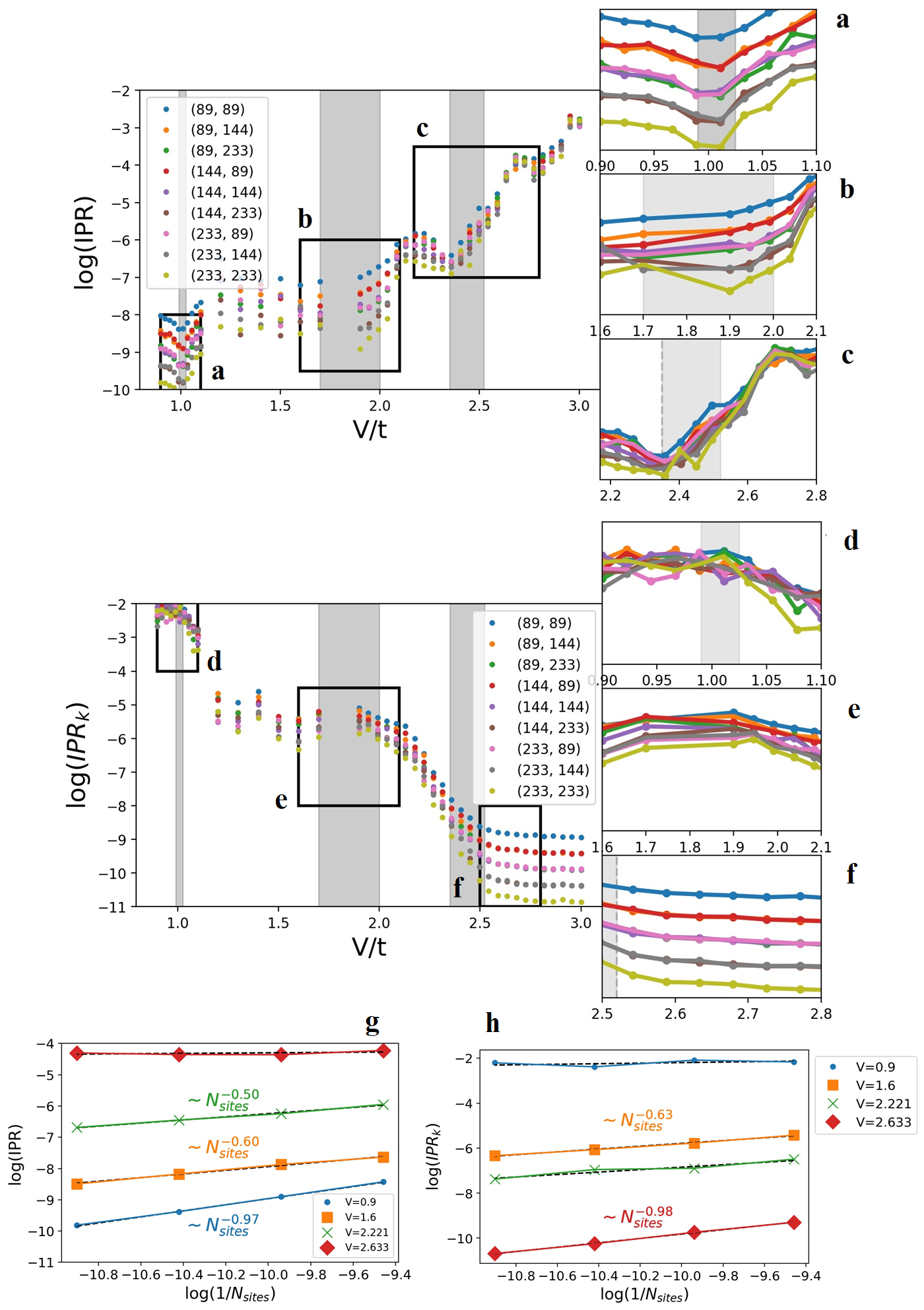}}\medskip{}
\par\end{centering}
\captionsetup{justification=raggedright,margin=0cm}
\caption{(a), (b), and (c) {[}(d), (e) and (f){]} show the value of log(IPR)
{[}${\rm log(IPR_{k})}${]} as a function of quasi-disorder strength
for different system sizes. In the inset axes are magnified areas
of interest to our study of the localization properties close to the
topological phase transitions (represented by the shaded regions).
(g,h) Scaling of the IPR and ${\rm IPR_{k}}$ for a few selected values
of quasi-disorder. \label{fig:IPR_square}}
\end{figure}
\par\end{center}

\subsubsection{Critical Chern Insulator - Anderson Insulator}

Increasing the incommensurate potential even further has evident effects
both in the spectral gap and in the localization of states. We can
see in Figure \ref{fig:gap-square}(b) that the data for the spectral
gap is very noisy presenting big fluctuations for all sizes, which
manifests itself in the big error bars in the values obtained through
thermodynamic-limit extrapolation of the gap. Nevertheless, we are
fairly certain that before the last topological phase transition,
still in the CCI phase, we have a, albeit very small, non-vanishing
spectral gap. The reason being that, since in that region we have,
undoubtedly, quantized Chern number, and the localization data in
Figs.$\,$\ref{fig:summary-square}(c),\ref{fig:IPR_square} indicates
that the eigenstates are critical around the gap edge, the only way
to conciliate these two facts is to have finite spectral gap.

The mechanism for the last topological phase transition, that happens
around the rightmost shaded area in Figs.$\,$\ref{fig:gap-square},\ref{fig:IPR_square},
is however less clear. We see in Fig.$\,$\ref{fig:gap-square}(b)
that the gap seemingly closes at the onset of this region and sharply
reopens before closing again. In this latter regime, the system clearly
becomes an AI: the eigenstates are localized due to the null value
of the Chern number and the convergence of the ${\rm IPR}$ into a
constant shown in Fig.$\,$\ref{fig:IPR_square}(c), which implies
that $\gamma\approx0$ and $\gamma_{k}\approx1$, as shown in Fig.$\,$\ref{fig:summary-square}(c).

\section{Conclusions}

\label{sec:conclusions}

In this paper, we explored the effects of quasi-disorder on Chern
insulators and found that it can lead to a rich and diverse landscape
of phase transitions. These go from transitions of the same nature
as the conventional transitions in clean systems to transitions with
no counterpart both in the clean and randomly disordered cases. 

Contrary to uncorrelated disorder, quasiperiodicity can induce gap
closing and reopening topological transitions. These can be of widely
different nature, including (i) trivial to Chern insulators, both
with ballistic states near the gap edges; (ii) Chern to Chern insulators
with critical states around the gap edges or (iii) Chern to trivial
insulators accompanied by a ballistic-to-localized transition for
the gap-edge states. While transition (i) has the same nature of clean-limit
topological transitions, transitions (ii) and (iii) have no clean-limit
counterpart. Additionally, we can have topological transitions for
which the gap closes and does not reopen. Among these, we found an
interesting transition from a Chern insulator to a trivial gapless
Anderson insulator, with intermediate metallic and critical phases
where the Chern number is not quantized. Curiously, we have not found
an instance of a gapless topological insulating phase (with localized
states at the Fermi level), which is common in systems with random
disorder.

Our findings demonstrate that quasi-disorder can induce a greater
variety of phase transitions than previously thought - much greater
than uncorrelated disorder - and that these transitions can be classified
into different categories based on their underlying physical mechanisms.
 Our results significantly expand the knowledge of the interplay between
quasiperiodicity effects and topology in Chern insulators, unveiling
a miriad of different phases and phase transitions in simple models
that can be experimentally realized in optical lattices \citep{Jotzu2014,Liu2018},
photonic \citep{PhysRevLett.100.013905,Wang2009,silveirinha2019_haldanePthotonic}
and acoustic \citep{PhysRevLett.122.014302,Ma2019} media, and possibly
even moiré systems \citep{HaldaneMoire}. 

\bibliographystyle{apsrev4-1}
\bibliography{refs}

\end{document}